\documentclass[letterpaper, 10 pt, conference]{ieeeconf}  

\usepackage{algorithm}
\usepackage{algpseudocode}
\usepackage{amsmath}
\usepackage{varwidth}
\usepackage{color}
\usepackage{url}
\usepackage{graphicx}
\usepackage{booktabs}
\usepackage{mathtools}
\usepackage{xcolor}
\usepackage{pgfplots}
\usepackage{float}

\pgfplotsset{compat=1.18}
\newtheorem{example}{Example}

\IEEEoverridecommandlockouts                              

\title{\LARGE \bf
Virtual Force-Based Routing of Modular Agents on a Graph
}

\author{Adam Casselman$^{1}$, Manav Vora$^{2}$, and Melkior Ornik$^{2}$ 
\thanks{*This work was supported by the Office of Naval Research grant N00014-23-1-2505.}
\thanks{$^{1}$ Department of Aerospace Engineering,
        Georgia Institute of Technology, Atlanta, GA, 30318, USA.
        {\tt\small acasselman3@gatech.edu}}%
\thanks{$^{2}$ Department of Aerospace Engineering and Coordinated Science Laboratory, University of Illinois Urbana-Champaign, Urbana, IL, 61801, USA.
        {\tt\small mkvora2@illinois.edu, mornik@illinois.edu}}%
}

\begin{document}

\maketitle
\thispagestyle{empty}
\pagestyle{empty}

 \begin{abstract}

Modular vehicles present a novel area of academic and industrial interest in the field of multi-agent systems. Modularity allows vehicles to connect and disconnect with each other mid-transit which provides a balance between efficiency and flexibility when solving complex and large scale tasks in urban or aerial transportation. This paper details a generalized scheme to route multiple modular agents on a graph to a predetermined set of target nodes. The objective is to visit all target nodes while incurring minimum resource expenditure. Agents that are joined together will incur the equivalent cost of a single agent, which is motivated by the logistical benefits of traffic reduction and increased fuel efficiency. To solve this problem, we introduce a novel algorithm that seeks to balance the optimality of the path that every single module takes and the cost benefit of joining modules. Our approach models the agents and targets as point charges, where the modules take the path of highest attractive force from its target node and neighboring agents. We validate our approach by simulating multiple modular agents along real-world transportation routes in the road network of Champaign-Urbana, Illinois, USA. The proposed method easily exceeds the available benchmarks and illustrates the benefits of modularity in multi-target planning problems.

\end{abstract}

\section{Introduction}

Modular systems are a recent framework within the umbrella of multi-agent systems, where agents have the ability to join and split from each other mid-mission. There is a growing interest in these reconfigurable technologies --- for example, supply chain companies are beginning to platoon trucks using communication between vehicles. Using a technique of \textit{drafting} which closely resembles the idea of modularity without the required connecting technology, the vehicles follow each other in close-proximity to reduce aerodynamic drag and travel 3-5 miles an hour faster \cite{7497531, liu2024alleviating}.  In general, modular vehicles benefit from flexibility and efficiency when they are joined while retaining the agility to perform different sub-tasks separately \cite{4141032}.

On the side of public transport, urban transit vehicles may benefit from modularity, lowering the operational cost of public transportation and increasing passenger convenience by better matching the frequency and size of vehicles to passenger needs at different stops \cite{CHENG2024104746}. If carefully designed, there additionally exists significant potential in passenger time-saving via \textit{enroute transfer}, by allowing passengers to choose an appropriate module in a joined vehicle before the individual modules split and head to different destinations \cite{filippi2025exploiting, 10422702}. 

\begin{figure}[H]
    \centering
    \includegraphics[width=0.9\linewidth]{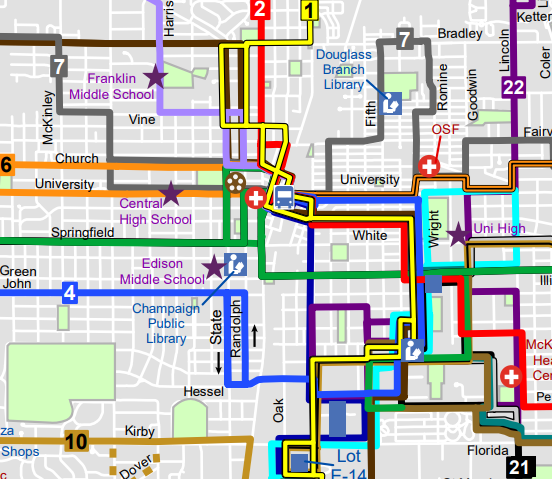}
    \caption{Local public transit authority routes along the central part of Champaign-Urbana, IL, USA \cite{mtd_schedule_2024}.}
    \label{fig:CUMTD}
\end{figure}

The motivating example discussed in this paper, previously introduced in \cite{jagdale2023optimal}, is a fleet of delivery vehicles that may share similar paths for a significant portion of their routes. Modules may dock to allow package transfer and leverage fuel efficiency benefits throughout their routes, simultaneously decreasing delivery transit time and operational expenditure. In our numerical examples, we look through the lens of routes localized in Champaign-Urbana, IL, USA. While delivery routes naturally vary by service and day, Fig.~\ref{fig:CUMTD} illustrates a closely related network of public buses in the central Champaign area. As bus services pass through the central part of the community, their paths often overlap for a significant distance before diverging to reach their final stops. These redundant paths suggest a potential benefit of joining during this period of time to take advantage of energy and logistical savings.


\subsection{Related work and contributions}

Abstracting away the domain details, routing of modular agents on a graph presents a significant challenge. Yet, previous work on routing modular agents is rather sparse. Developed models propose a heuristic approach based on a nearest neighbor cost estimation and determination of joining/splitting by leveraging centrality of nodes between remaining destinations and agents' position on the road network. To the best of our knowledge, the method from Jagdale and Ornik \cite{jagdale2023optimal} is the only known routing algorithm for a general problem of modular multi-target routing, where modules need to visit a predetermined set of targets. \textit{However, the framework of \cite{jagdale2023optimal} is designed for, and the workings of its heuristic method limited to, two modules only.} Other algorithms \cite{filippi2025exploiting, 10422702, khan2023application} explore modularity within the context of a bus transit, largely considering strategies to split and join based on passenger demand and headway threshold.

This paper proposes an algorithm that routes an arbitrary number of modules to predetermined set of target nodes. The algorithm simultaneously routes agents and determines joining or splitting action based on the formulation of a \textit{discretized potential field model}. In this approach, agents are attracted to their assigned targets, but also experience attraction in the direction of nearby agents. Intuitively speaking, agents that are far away from their respective targets will experience a higher attractive forces towards each other and will join until they are close enough to their targets at which point it is cost-effective to split. One apparent challenge to this methodology is that potential fields are typically implemented in continuous space and time. Routing on a discrete state space like a graph requires a nuanced approach to discretization of such a physical phenomenon. 

While \cite{jagdale2023optimal} showed that the modular agent routing problem is a special case of the traveling salesman problem, and it is thus NP-hard to obtain an optimal solution, we validate our approach using batch simulations of agent sets ranging from $n=2$ to $n=20$ agents on a real-world map, showing that our method significantly outperforms both previous work \cite{jagdale2023optimal} and a natural non-modular benchmark.



In the remainder of the paper, Section II outlines the modular agent routing problem, and Section III derives our approximate solution using a force-based algorithm attract agents to a set of targets. We then discuss the computational expenditure in Section IV. Lastly, we demonstrate our results through numerical examples and determine performance compared to existing benchmarks in Section V.

\section{Problem Statement}

This paper's formal setup considers the optimal route planning of modular agents traversing an edge-weighted directed graph $\mathcal{G(V,E)}$, where nodes $\mathcal{V}$ are connected by edges $\mathcal{E}$. The mathematical definition for this problem is identical to the one stated, \textit{but not solved for an arbitrary number of agents}, in \cite{jagdale2023optimal}. Agents are required to reach a set of targets $\mathcal{T}\subset\mathcal{V}$ while minimizing the total cost of travel. Agent modularity allows the modules to join and split at any node in the graph, where joined modules --- now acting as a single agent --- only incur the edge traversal cost of a single module until they later split.  Lastly, a module can choose to not move if there is a cost benefit to do so; in other words, $\mathcal{E}$ includes self-loops at every node. For example, if there are two modules that share the same path but are separated by several nodes, one may wait for the other. 


We now formally introduce the problem formulation. Modules make their way to targets through a series of nodes connected by graph edges. When a module traverses along an edge $e \in \mathcal{E}$, it incurs a cost according to the edge weight $w_e > 0$. In a practical sense, this weight can be interpreted as energy expenditure between two geographical points of interest. As in \cite{jagdale2023optimal}, we assume that traversal times along each edge are the same; this assumption can be trivially relaxed by partitioning each edge into smaller pieces, with the traversal time for each piece equal to the smallest common denominator of original traversal times.

We define the set of all edges traversed by at least one module at time $t$ by $\mathcal{K}(t)$. A mission ends when all target nodes $\mathcal{T}$ have been reached by at least one agent; we refer to the time step at which the mission is completed by $T$. The resultant cost incurred by all modules is
\begin{equation}
\sum_{t=1}^{T} \sum_{e\in\mathcal{K}(t)} w_e.    
\label{cost_fnc}
\end{equation}
The problem of optimal planning is thus given as follows.



\textit{Problem 1:} Let $n$ modules operate on a graph $\mathcal{G(V,E)}$
 with a target set $\mathcal{T} \subset
\mathcal{V}$. Denote the path of module \textit{i} by
 $P_i = (v_i(0),\ldots, v_i(T))$ with $v_i(t) \in \mathcal{V}$ and $(v_i(t), v_i(t +
 1)) \in \mathcal{E}$ for each $0 \leq t < T$. Assume that $v_1(0),\ldots,v_n(0)$ are fixed. Determine 
 \begin{equation*}
 \begin{split} \mathop{\mathrm{argmin}}_{P_1,\ldots P_n} & \quad \sum_{t=1}^{T} \sum_{e\in\mathcal{K}(t)} w_e \\
 \textrm{such that} & \quad \mathcal{T} \subset \bigcup^n_{i=1} \bigcup^T_{t=0} \{v_i(t)\}\textrm{.}
 \end{split}
 \end{equation*}

\section{Force-Based Algorithm}

We now present an overview of the algorithm. Problem~1 is NP-hard in terms of the number of agents operating on the graph as shown in \cite{jagdale2023optimal}, and therefore an analytical solution is not computationally feasible. Instead, we desire a heuristic that finds an approximately optimal solution. We use a unified mechanism motivated by a \textit{point charge representation} of the system to determine \textit{agent-target} and \textit{agent-agent attractions}, and use those for optimal planning. 

Both classes of attractions are modeled using an inverse-squared distance law \cite{voudoukis2017inverse}, where the distance is derived from the sum of edge weights along several possible paths that a module would take a target or other agents. 
At each timestep, agents reevaluate the force contributions along all adjacent edges, and select to traverse the one with the maximum net force. At every timestep this process is repeated, allowing for dynamic re-evaluation as the agent states evolve and targets are visited. We discuss these steps in detail below.



\subsection{Target Assignment}

Our algorithm begins by assigning the next target to each agent. We do so by using Dijkstra's Algorithm to find the shortest path from each agent's location to the set of unvisited targets and implement a standard nearest neighbor method \cite{dhakal2008hybrid,walsh2018agent}, where each agent is assigned the target that minimizes its individual path cost. While this assignment scheme does not account for possible conflicts between agents' assignments, it serves as a computationally efficient policy that is typical in classical path planning literature \cite{cap2013multi}.  

\subsection{Path Sampling}

In order to benefit from the cost-savings of modularity, it might not be optimal for each agent to follow a shortest path to its assigned target. Thus, we first determine a set of \textit{candidate edges} for each agent to traverse. We utilize Yen's Algorithm \cite{yen1971kshortest} to determine the $k$ shortest loopless paths from agent $i$'s current position $s_{i,t}$ to its assigned target node $g_{i,t}$ on the graph $\mathcal{G}$, as well as to the other agents' current positions $s_{j,t}$. In practice, we choose $k$ empirically based on the connectivity of the graph since agents will benefit from a broader range of candidate edges if there are more edges connecting each node. As opposed to a single shortest path, a larger sample size enables agents to evaluate several meaningful directions through the graph. 

In terms of notation, let \( \mathcal{P}_{i,j}^{(k,t)} = \{ s_{i,t}\xlongrightarrow{1}s_{j,t}, s_{i,t}\xlongrightarrow{2}s_{j,t}, \ldots, s_{i,t}\xlongrightarrow{k}s_{j,t} \} \) denote the set of the \( k \) shortest paths from \( s_{i,t} \) to \( s_{j,t} \) when $i\neq j$, where each path is denoted by $s_{i,t}\xlongrightarrow{l}s_{j,t}$. Analogously, let \( \mathcal{P}_{i,i}^{(k,t)} = \{ s_{i,t}\xlongrightarrow{1}g_{i,t}, s_{i,t}\xlongrightarrow{2}g_{i,t}, \ldots, s_{i,t}\xlongrightarrow{k}g_{i,t} \} \). Each path is an ordered sequence of edges from origin to destination; we denote the $r$-th edge in the $l$-th path of $\mathcal{P}_{i,j}^{(k,t)}$ by $e_{i,j,l}^r\in\mathcal{E}$, dropping the dependence on $t$ for notational simplicity. The candidate edges are then all edges $e_{i,j,l}^1$, where we note that some of these edges might coincide.


\subsection{Force Computation}


Agent $i$'s next traversed edge is selected among edges $e_{i,j,l}^1$ by using the notion of \textit{attractive forces}. Inspired by the inverse-square law \cite{voudoukis2017inverse}, these forces are computed using the weighted sum of the edges along each of the $k$ lowest weight paths connecting two nodes on the graph. Along each path $s_{i,t}\xlongrightarrow{l}s_{j,t}$ or $s_{i,t}\xlongrightarrow{l}g_{i,t}$, we compute an \textit{attractive force} based on the total weight of edges along that path. Namely, if $d_{i,j,l}$ is the total weight of edges along path $s_{i,t}\xlongrightarrow{l}s_{j,t}$ (or $s_{i,t}\xlongrightarrow{l}g_{i,t}$ for $i=j$), then we define \begin{equation}
F_{i,j,l}^{\text{att}} = \frac{\alpha_j}{d_{i,j,l}^2}\textrm{,} 
\label{att}
\end{equation}
where $\alpha_j$ is a tunable scaling constant. 

We generally take all $\alpha_j$ for $j\neq i$ to be constant and equal $\alpha$, and allow $\alpha_i$ to be a different constant $\beta$. The relative tuning between $\alpha$ and $\beta$ governs agent tendency to bias towards more independent or collective formations. For example, if $\alpha=0$ we effectively remove module-module attraction, so the system reduces to a variation of the classical multi-agent path finding problem \cite{stern2019multi}. 

For each candidate path, we extract the first edge  $e_{i,j,l}^1$ as the \textit{direction of influence of its attractive force}. The total attractive force affecting module $a$ along a candidate edge $\hat{e}\in\mathcal{E}$ is then given by 

\begin{equation}
\vec{f}_{i,\hat{e}} = \sum_j\max_{1\leq l\leq k}\{F_{i,j,l}^{att}~|~e_{i,j,l}^1=\hat{e}\}\textrm{,}
\label{att2}
\end{equation}
where the $\max$ inside the sum is taken to be $0$ if the set $\{F_{i,j,l}^{att}~|~e_{i,j,l}^1=\hat{e}\}$ is empty, i.e., if none of the candidate paths $s_{i,t}\xlongrightarrow{l}s_{j,t}$ start by traversing $\hat{e}$. module $i$ then traverses edge $\hat{e}$ which maximizes \eqref{att2}.

By using $\max$ in \eqref{att2}, the algorithm only considers the shortest path $s_{i,t}\to s_{j,t}$ (or $s_{i,t}\to g_{i,t}$) that passes through $\hat{e}$. A natural modification of \eqref{att2} would be to consider the sum of forces induced by all such paths; in practice, such a change might not make a large difference, especially when one path is much shorter than the rest. 

\begin{example}

\begin{figure}[H]
    \centering
    \includegraphics[width=0.8\linewidth]{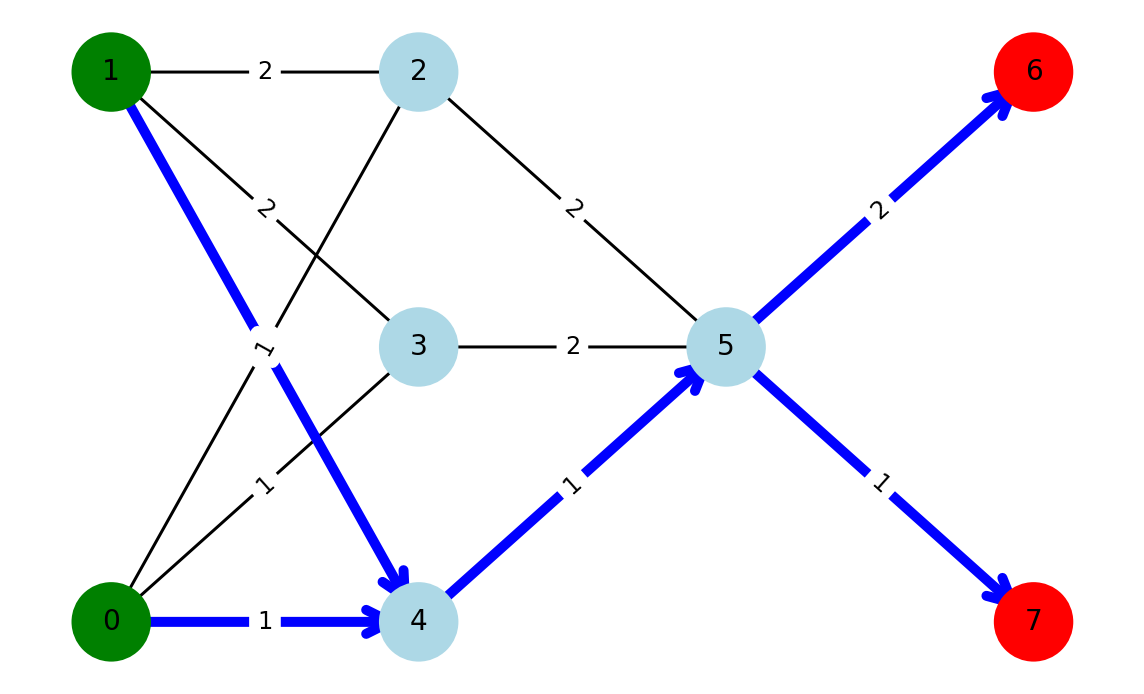}
    \caption{A visualization of node-to-node force-based computation. The modules start at node 0 and node 1 denoted in green, and the targets are node 6 and node 7 denoted in red; the weights of each edge are denoted above the edge.}
    \label{fig:force_comp_ex}
\end{figure}

Fig. \ref{fig:force_comp_ex} illustrates the proposed assignment of attractive forces with two modules and two targets. Modules $a_0$ and $a_1$ start at nodes 0 and 1, whose respective targets are chosen to be nodes 6 and 7. With $\alpha=\beta=1$ and $k=3$, we first consider three shortest paths from agent $a_0$ to the assigned target node 6: $(0,4,5,6)$, of total cost $4$, $(0,3,5,6)$, of total cost $5$, and $(0,2,5,6)$, of total cost $5$. We also consider the three shortest paths from $a_0$ to module $a_1$: $(0,4,1)$, of total cost $2$, and $(0,3,1)$, and $(0,2,1)$, both of total cost $3$. The force induced along edge $(0,4)$ is thus $1/2^2+1/4^2$, which dominates the forces along edges $(0,3)$ and $(0,2)$ equaling $1/3^2+1/5^2$. Thus, module $a_0$ moves to node $4$ in its first move. Analogously, module $a_1$ will move to node $4$ in its first move, after which the joint agent proceeds with the next step in the mission.
\end{example}

\subsection{Moving-or-Waiting Logic}
As described so far, the proposed strategy might result in modules continually oscillating between a small subset of nodes. A representative example occurs if two agents, $A$ and $B$, are separated by a single edge but are relatively far away from their respective targets. In this case, the agent-agent attraction may dominate the agent-target interaction leading to a state in which the edge with the highest force corresponding to module $A$ is the current position of $B$ and vice versa. 

To address this deadlock, we implement a waiting policy allowing agents to delay their movement based on the state of the system. In particular, we consider situations where two agents, are intending to move to the others' respective states. In that case, the agent whose path to the assigned target is shorter is ordered to wait for one timestep for the other agent to join, or at least change position. If two agents have the same cost to the assigned target, one of the two is randomly ordered to wait.

The force computation and decision-making strategy for each agent at time step $t$ is shown in \textbf{Algorithm 1} below. 

\begin{algorithm}[!h]
  \caption{Agent Force Computation and Action Selection}
  \begin{algorithmic}[1]
    \While{not all targets have been visited}
      \For{each agent \(i\)}
          \State Assign agent to next nearest target $g_i$
        \State
          Compute \(k\) shortest paths to \(g_i\) 
        \State
            Compute \(k\) shortest paths to all agents \(j \neq i\)
      \vskip 2pt
        \State
        \begin{varwidth}{0.8\linewidth}
            Apply \eqref{att} to obtain individual forces $F_{i,j,l}^{\textrm{att}}$ for all computed paths
        \end{varwidth}
        \State
        \begin{varwidth}{0.8\linewidth}
            Apply \eqref{att2} to obtain total forces $\vec{f}_{i,\hat{e}}$ for all candidate edges
        \end{varwidth}
      \vskip 2pt
      \EndFor
      \vskip 3pt
      \State
        \begin{varwidth}{0.8\linewidth}
            Move all agents $i$ simultaneously along edges \(\hat e_i^*\) which maximize $\vec{f}_{i,\hat{e}}$ or wait if paths overlap
        \end{varwidth}
        \vskip 2pt

    \EndWhile
  \end{algorithmic}
\end{algorithm}

\begin{example}
\begin{figure}[H]
    \centering
    \includegraphics[width=0.7\linewidth]{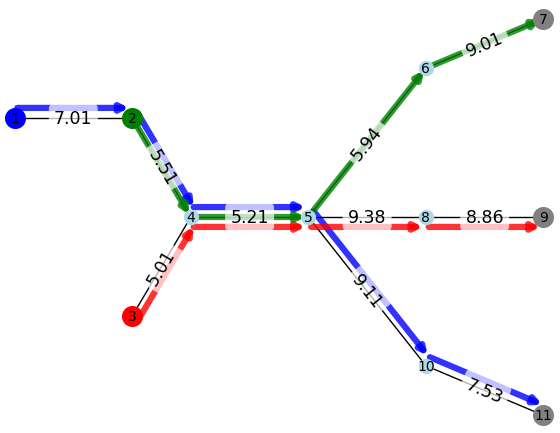}
    \caption{A toy graph showing the benefits of delaying movement. There are three agents on this graph, whose paths are represented by red, blue, and green edges. Their targets are denoted by gray nodes.}
    \label{fig:3_agent_wait}
\end{figure}

We demonstrate the benefits in delaying movement with a brief illustrative scenario shown in Fig.~\ref{fig:3_agent_wait}, used in \cite{jagdale2023optimal}. Three modules initialize at nodes $\{1,2,3\}$ and must reach their respective targets $\mathcal{T}=\{7,9,11\}$. Algorithm 1 would produce paths $(1,2,4,5,10,11)$, $(3,4,4,5,8,9)$, and $(2,2,4,5,6,7)$ for the three agents respectively. Edge $(2,4)$ is shared by the blue and green agent. The green module waits at node $2$ for the blue module where they immediately join. At the next timestep, the red module waits for these two modules at node $4$. Thus, edge $(4,5)$ is shared by all three modules. The cost incurred by this system is $76.49$. If these modules did not have the ability to wait, they would not experience the same cost savings, and would take paths $(1,2,4,5,10,11)$, $(3,4,5,8,9)$, and $(2,4,5,6,7)$.
\end{example}

\section{Complexity Analysis}
\label{compan}

While paying the price of suboptimality, our heuristic algorithm is significantly less computationally costly than the optimal solution, proven to be NP-hard \cite{jagdale2023optimal}. To show this difference, we briefly analyze the complexity performance of Algorithm~1.

Let $n$ denote the number of modules, $m=|\mathcal{V}|$ the number of nodes, and  $E=|\mathcal{E}|\leq m^2$ the number of edges on the graph $\mathcal{G}(\mathcal{V,E})$. Lines 3--5 of Algorithm 1 employ Dijkstra's and Yen's Algorithm to assign targets and generate the $k$ shortest loopless paths between two nodes, with the time complexity of $O(k\cdot(m\log(m)+E))$ for each considered pair of nodes \cite{yen1971kshortest}. There are up to $n^2$ pairs of nodes which contain modules, and no more than $nm$ module-target pairs. The complexity of nearest-neighbor assignment can obviously be included within the above higher complexity of Dijkstra's and Yen's Algorithm.

Once the candidate paths are populated, we then compute the force-based interactions. Module-target forces are evaluated across each of the $k$ paths resulting in complexity $O(km)$ for each agent and timestep, while module-to-module interactions are pairwise, resulting in a total of $O(kmn)$ computations per agent and time step. Empirically, we found that $k$ should be a fixed constant and relatively small in practice, as grid-like road networks by definition translate into sparse graphs with relatively uniform topology. Thus, the total complexity of computing force-based interactions is $O(nm + n^2m)$ per time step. Finally, the determination whether agents should wait depends on whether their planned paths overlap; since each path is of length no larger than $m$, and there are no more than $n^2$ pairs of agents, the complexity is again $O(n^2m)$.

The computations above are performed for $T$ time steps until targets are reached. Given that a path between two targets cannot follow a loop, the worst case requires every agent to take $m$ steps between every two targets, leading to a total of $T\leq m^2$ steps. The total time complexity thus computes to 
\begin{equation*}
\begin{split}
& O(m^2((m\log(m)+m^2)(mn+n^2) + mn + n^2m)) = \\
& O(m^2(m^3n+m^2n^2))=O(m^5n+m^4n^2)\textrm{.}
\end{split}
\end{equation*} For comparison, the method in \cite{jagdale2023optimal}, limited to $n=2$, reports a complexity of $O(m^8)$; for $n=2$ our method yields $O(m^5)$.

\section{Numerical Results}

To illustrate the behavior induced by the proposed algorithm, we first provide an example describing the policies of three modules on a multi-target mission in Champaign, IL, USA. To evaluate the effectiveness of the proposed algorithm, we then conduct a series of batch experiments. Namely, we compare our approach against the heuristic method for modular agents used by Jagdale and Ornik \cite{jagdale2023optimal} for the two-module case, and in the absence of known benchmarks for scenarios of more than two modules, we also implement a \textit{non-modular baseline} where agents do not have the ability to join or split. The results show that our algorithm outperforms both the heuristic in \cite{jagdale2023optimal} for $n=2$ and the costs of non-modular agents for $n\geq 2$. We additionally provide a short study of the sensitivity of the proposed algorithm to changes in tuning parameters, indicating the best values for agent-agent and agent-target scaling constants $\alpha$ and $\beta$.

\subsection{Illustrative Example: Champaign, IL}

\begin{figure}[H]
    \centering
    \includegraphics[width=0.9\linewidth]{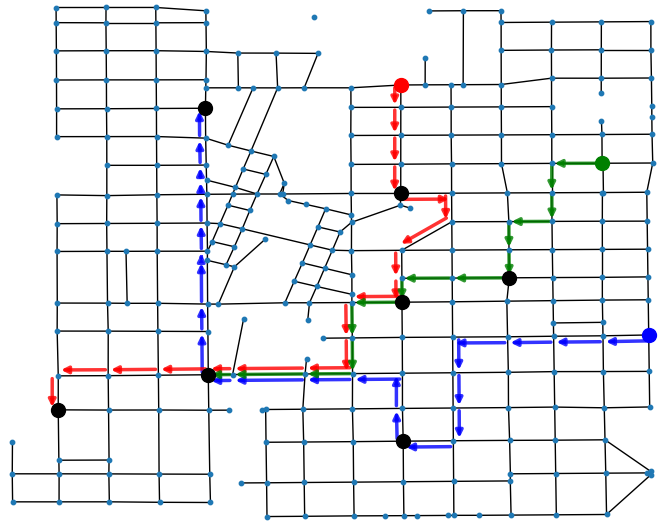}
    \caption{A graph of a multi-target routing task on a map of Champaign, IL, USA. Three modules with paths denoted in red, blue, and green start at their respective colors and reach points of interest around the city denoted by black nodes.}
    \label{fig:3_agent_CU}
\end{figure}

We investigate the potential benefits of using modular vehicles to facilitate movement through the city of Champaign, IL, USA, using \cite{graphml} to create a realistic map. The agents start at three existing locations, and must reach several points of interest, including academic buildings, bookstores, student housing, grocery stores, and restaurants which are all denoted by black nodes. Tuning parameters $\alpha$ and $\beta$ are set to $\alpha=0.5$ and $\beta=1$, in line with sensitivity results described later in Section~\ref{sens}. We fix the number of sampled shortest paths to $k=5$. Fig.~\ref{fig:3_agent_CU} illustrates the agent behavior resulting from applying the proposed routing algorithm.

Initially, the modules are spread out across the eastern part of the map; they reach the outer, easternmost targets first and work their way west and south. The red and green modules arrive at their targets first and start to experience significant agent-agent towards each other. They convene at a common node in a concentrated area of downtown Champaign, and jointly move over several blocks. The blue module eventually joins them as well, and all three modules feels move along a series of common edges. 

Nearing the end of the task, the modules have no benefit of remaining joined and the size of agent-target attraction takes over; the blue module gets assigned one target, the red gets assigned another, and with no further targets available, the green module terminates movement.

We observe that modules, as expected, joined in an area with a denser set of targets. Because these areas have more modules traveling through them, they are prime locations to take slightly suboptimal individual routes to benefit from joining. The modules then split in areas where targets are few and far between, and remaining joined would incur a significant penalty in suboptimal routing.

\subsection{Two-Module Benchmarking}




For stronger validation of the proposed algorithm, we first consider the two-module delivery scenario from \cite{jagdale2023optimal}. We note that the example shown in Fig.~\ref{fig:3_agent_wait} appears in \cite{jagdale2023optimal} as well, and the agent behavior of the two algorithms --- ours and that of \cite{jagdale2023optimal} --- is identical in that scenario.

Going beyond single examples, the work in this section considers routing agents engaged on $100$ multi-target missions on a graph of central Champaign, IL, USA described in Fig.~\ref{fig:3_agent_CU} and derived from \cite{graphml}. Each case contained 10 targets, where our parameters are $\beta = 2\alpha=1$ and $k=5$. We compare the proposed algorithm with two different routing methods, one devised by Jagdale and Ornik \cite{jagdale2023optimal} and the other which considers a pair of \textit{non-modular vehicles} which follow a standard nearest-neighbor algorithm \cite{dhakal2008hybrid}. Table \ref{tab:routing_optimality_2} shows the frequency at which each of these methods resulted in the lowest-cost routing among the three. We note that the sum of frequencies exceeds $100\%$, notably because the routes generated by \cite{jagdale2023optimal} and our algorithm often coincide. However, our method in general outperforms the others, equaling or exceeding the performance of the other two in 97\% of the simulations.

\begin{table}[h]
\centering
\caption{Frequency of lowest-cost routing for two agents}
\resizebox{.425\textwidth}{!}{%
\begin{tabular}{l c}
\hline
\textbf{Routing Method} & \textbf{Percentage of Times Best} \\
\hline
Non-modular & 3\% \\
Method in \cite{jagdale2023optimal} & 68\% \\
Force-based \textit{(ours)} & 97\% \\
\hline
\end{tabular}%
}
\label{tab:routing_optimality_2}
\end{table}

\subsection{Batch Simulation on the Champaign-Urbana graph}

The algorithm in \cite{jagdale2023optimal} does not apply to more than two modules and, to the best of the authors' knowledge, there do not exist any other algorithms for multi-target routing of modular agents. We thus compare to a non-modular baseline for $n>2$ modules. 

We continue using the same graph of Champaign, IL as in the previous sections for this set of experiments, and  again conduct $100$ missions of multi-target reachability. In each mission, $n$ modules are randomly assigned starting nodes corresponding to existing bus stops, and randomly assigned a set of $2n$ targets on the graph.

\begin{figure}[h]
    \centering
    \includegraphics[width=0.9\linewidth]{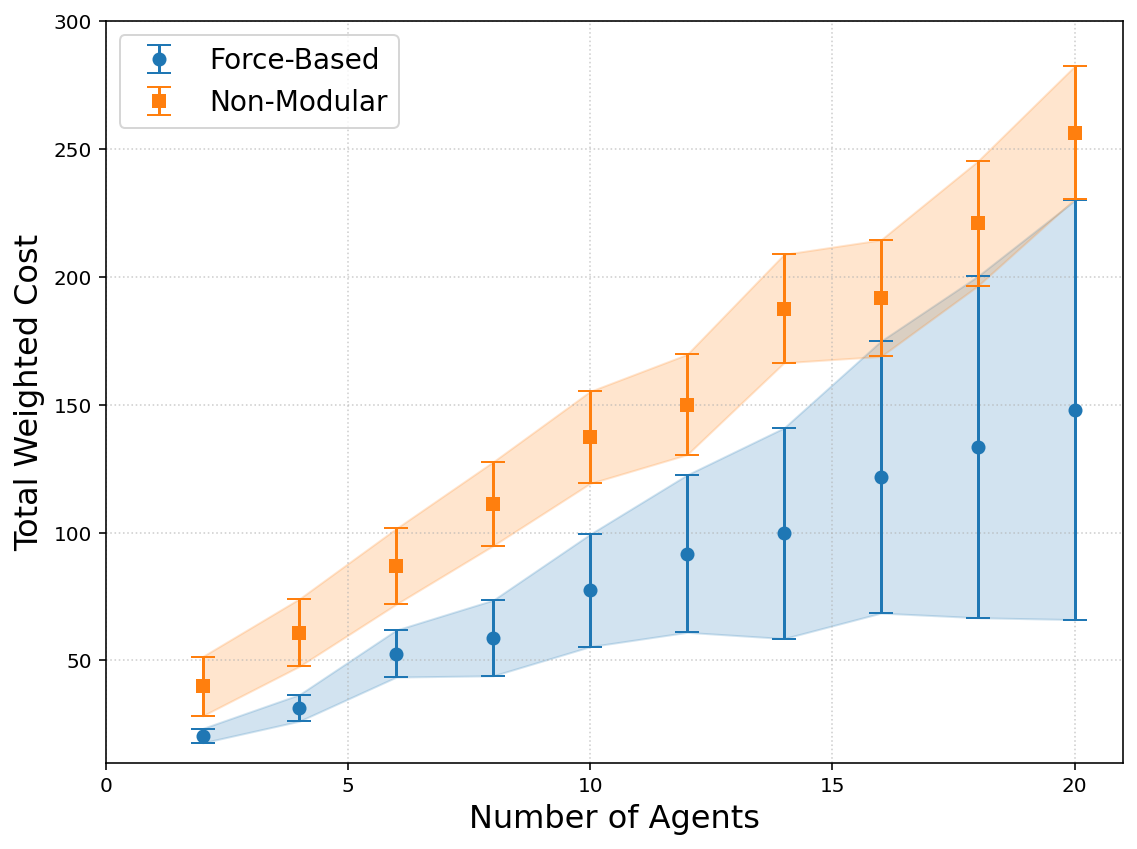}
    \caption{Cumulative path cost as a function of module number $n$. The force-based algorithm consistently maintains lower global cost compared to the baseline, albeit at a price of higher variance for a larger number of agents. The markers indicate the mean cost over 100 trials, and the vertical lines show the variance of cost through the trials.}
    \label{fig:mean_variance}
\end{figure}

The performance obtained using empirically chosen parameters $\alpha = 0.5$, $\beta = 1$, and $k=5$ is shown in Figure~\ref{fig:mean_variance}.


The proposed force-based method is consistently more efficient than the non-modular baseline across the sampled populations, with the largest difference for a small number of agents. As more agents are introduced, the variance for the force-based method becomes much larger, with modules sometimes choosing to incur a high cost in an effort to join with others. Nonetheless, even in such scenarios the force-based modular algorithm almost invariably outperforms the baseline. 

\subsection{Sensitivity Analysis}
\label{sens}

Performance of the proposed algorithm depends on the choices of parameters $\alpha$ and $\beta$ which scale the agent-agent and agent-target forces respectively. Using a set of pseudorandom initial configurations for $n=5$ agents and a random set of $2n=10$ targets across the Champaign graph used throughout this section, we performed stochastic simulations for each parameter pair and conducted a sensitivity analysis to characterize the tradeoff between the scaling of these two objectives. We characterize the optimality of the force-based heuristic in Fig.~\ref{test}, where the color gradient represents changes in observed performance for different choices of parameter pairs. As the algorithm performance only depends on the quotient of $\alpha$ and $\beta$ and not their individual values, we naturally scaled to $\alpha,\beta\in[0,1]$. Values on the heatmap closer to $1$ represent the sets of parameters that produced the lowest mean cost for the $100$ trials. 

\begin{figure}[H]
\begin{center}
\begin{tikzpicture}
\begin{axis}[
    width=7cm,
    height=7cm,
    xlabel={$\beta$},
    ylabel={$\alpha$},
    xmin=0, xmax=1,
    ymin=0, ymax=1,
    xtick={0.1,0.3,0.5,0.7,0.9},
    xticklabels={0,0.25,0.5,0.75,1},
    ytick={0.1,0.3,0.5,0.7,0.9},
    yticklabels={0,0.25,0.5,0.75,1},
    colorbar,
    colormap/viridis,
    point meta min=0,
    point meta max=1,
    enlargelimits=false,
    axis on top,
    grid=none,
    view={0}{90},
]

\addplot [
    matrix plot*,
    point meta=explicit,
    mesh/cols=5,
] table [meta=z] {
x y z
0.1 0.1 0.012
0.3 0.1 0.312
0.5 0.1 0.312
0.7 0.1 0.312
0.9 0.1 0.312

0.1 0.3 0.00
0.3 0.3 0.76
0.5 0.3 1.0
0.7 0.3 0.815
0.9 0.3 0.70

0.1 0.5 0.00
0.3 0.5 0.50
0.5 0.5 0.760
0.7 0.5 0.40
0.9 0.5 1

0.1 0.7 0.00
0.3 0.7 0.60
0.5 0.7 0.57
0.7 0.7 0.76
0.9 0.7 0.90

0.1 0.9 0.00
0.3 0.9 0.175
0.5 0.9 0.43
0.7 0.9 0.55
0.9 0.9 0.76
};

\end{axis}
\end{tikzpicture}
\end{center}

\caption{Heatmap of algorithm performance with respect to the choices of tuning parameters $\alpha$ and $\beta$.}
\label{test}
\end{figure}

The pairs with the highest heatmap value --- i.e., the pairs that indicate regions of highest performance that successfully balance individual goal-seeking with edge sharing coordination --- appear to be those where $\beta\approx 2\alpha$. This observation motivates our choice of $\beta=1=2\alpha$ throughout this section. 

While the number of sampled shortest paths $k$ is also a tunable parameter, we observed --- as mentioned in Section~\ref{compan} --- that the performance of the proposed algorithm in structured grid environments does not significantly increase with an increase in $k$. We thus maintain a low $k=5$ throughout the simulations in this section, leaving the algorithm computationally tractable.

\section{Conclusion}

This paper considers the problem of optimal routing for the developing framework of modular vehicles, where vehicles can choose to join or split mid-task, incurring resource cost savings when they are joined, but possibly paying the price of suboptimal routing if going out of their way to join with another vehicle. Focusing on multi-target tasks, we note that prior work established that the optimal solution to such a problem is not computationally feasible. However, no approaches for more than two modules have been proposed prior to this paper.

We propose a routing strategy based on a combination of nearest-neighbor routing and virtual attraction forces between vehicles. Our strategy is applicable to an arbitrary number of vehicles, computationally scales to support real-world, and significantly outperforms both the only existing benchmark --- which is limited to only two vehicles --- and a natural naive strategy for larger systems.

Despite its initial success outlined in this paper, the proposed method leaves room for improvement. As one direction, we note that the current method of target reassignment --- a multi-agent adaptation of nearest-neighbor policy --- is efficient over short-time horizons but fails to consider long-term optimality of the system. A possible alternative could include a learning-based target assignment procedure to prioritize long-term optimality. 

Furthermore, the proposed work, focused solely on reaching all targets by \textit{any} vehicle, does not address the existence of specific origin-destination pairs which underlies public transportation systems. On a similar note, the approach communicated in this paper optimizes over static graphs, and does not consider robustness to changed traffic density, vehicle availability or target demand. Future iterations should be able to re-plan based on real-time information about the operating graph and the vehicles' joint task.

\addtolength{\textheight}{-12cm}   

\bibliographystyle{ieeetr}
\bibliography{citations}

\end{document}